\documentclass[11pt,letterpaper,singlespacing, abstract=on]{scrartcl}
\usepackage[utf8]{inputenc}
\usepackage[usenames,dvipsnames]{xcolor}
\usepackage{amsmath,amssymb, amsfonts, amsthm, mathtools} 
\usepackage{tikz}
\usepackage{subfigure}
\usepackage{microtype}
\usepackage{paralist}
\usepackage{stmaryrd}
\usepackage{fancyhdr}
\usepackage{tabularx}
\usepackage[margin=1in]{geometry}
\usepackage{setspace}
\usepackage{array}
\usepackage{titlesec}
\usepackage{parskip}
\usepackage[pdfpagelabels=true]{hyperref}
\usepackage{listings}
\usepackage[T1]{fontenc}
\usepackage{etoolbox}
\usepackage{standalone}

\usetikzlibrary{decorations.pathreplacing}

\definecolor{darkgreen}{rgb}{0,0.6,0}

\def\Nscr{\mathcal{N}}
\def\Bscr{\mathcal{B}}

\def\Lscr{\mathcal{L}}

\def\cupp{\stackrel{.}{\cup}}

\renewcommand{\mid}{:}

\renewcommand{\epsilon}{\varepsilon}

\newtheoremstyle{mytheorem}{0.5cm}{0.5cm}{\itshape}{}
   {\bfseries}{}{\newline }{#1\, #2\, #3}
\theoremstyle{mytheorem}

\newtheoremstyle{note}{0.5cm}{0.5cm}{\upshape}{}
   {\bfseries}{}{\newline}{#1\, #2\, #3}
\theoremstyle{note}

\theoremstyle{mytheorem}
\newtheorem{theorem}{Theorem}

\theoremstyle{mytheorem}
\newtheorem{proposition}[theorem]{Proposition}

\theoremstyle{mycorollary}

\theoremstyle{mytheorem}
\newtheorem{lemma}[theorem]{Lemma}

\theoremstyle{mytheorem}

\theoremstyle{note}

\theoremstyle{note}

\def\cupp{\stackrel{.}{\cup}}

\begin{document}

\title{Approaching \boldmath $\frac{3}{2}$ for the $s$-$t$-path TSP}
\author{Vera Traub \and Jens Vygen}
\date{\small Research Institute for Discrete Mathematics and Hausdorff Center for Mathematics,\\
University of Bonn \\
\texttt{\{traub,vygen\}@or.uni-bonn.de}}

\begingroup
\makeatletter
\let\@fnsymbol\@arabic
\maketitle
\endgroup

\begin{abstract}
We show that there is a polynomial-time algorithm with approximation guarantee $\frac{3}{2}+\epsilon$
for the $s$-$t$-path TSP, for any fixed $\epsilon>0$.

It is well known that Wolsey's analysis of Christofides' algorithm also works for the $s$-$t$-path TSP
with its natural LP relaxation except for the \emph{narrow cuts} 
(in which the LP solution has value less than two). 
A fixed optimum tour has either a single edge in a narrow cut 
(then call the edge and the cut \emph{lonely}) or at least three
(then call the cut \emph{busy}).
Our algorithm ``guesses'' (by dynamic programming) lonely cuts and edges.
Then we partition the instance into smaller instances and strengthen the LP, 
requiring value at least three for busy cuts. 
By setting up a $k$-stage recursive dynamic program, we can compute a spanning tree $(V,S)$ and an LP solution $y$
such that $(\frac{1}{2}+O(2^{-k}))y$ is in the $T$-join polyhedron, 
where $T$ is the set of vertices whose degree in $S$ has the wrong parity.
\end{abstract}

\section{Introduction}

The traveling salesman problem has played a crucial role in combinatorial optimization for many decades.
Despite a lot of research, Christofides' classical $\frac{3}{2}$-approximation algorithm from 1976 is still unbeaten.
However, this ratio holds only if the tour is to begin and end in the same point.
For the more general problem where the given endpoints of the tour can be distinct (the $s$-$t$-path TSP),
 the classical algorithm achieves only the approximation ratio $\frac{5}{3}$.
Initiated by the work of An, Kleinberg, and Shmoys \cite{AnKS15}, there has been much progress recently, obtaining better
and better approximation ratios and introducing various interesting techniques. 
In this paper we obtain an approximation ratio arbitrarily close to $\frac{3}{2}$ by a completely new approach.

An instance of the $s$-$t$-path TSP consists of a finite metric space $(V,c)$ and $s,t\in V$. 
The goal is to compute a path $(V,H)$ with endpoints $s$ and $t$ (or a circuit if $s=t$) 
that contains all elements of $V$.
Christofides \cite{Chr76}, Serdjukov \cite{Ser78}, and Hoogeveen \cite{Hoo91} proposed to compute a cheapest spanning tree $(V,S)$, let 
$T:=\{v\in V:|S\cap \delta(v)| \text{ odd}\}\triangle\{s\}\triangle\{t\}$ be the set of 
vertices with wrong parity,
compute a perfect matching $M$ on $T$ and an Eulerian trail from $s$ to $t$ in $(V,S\cupp M)$,
and shortcut whenever a vertex is visited more than once.
This algorithm has approximation ratio $\frac{3}{2}$ for $s=t$ \cite{Chr76}, 
but only $\frac{5}{3}$ for $s\not=t$ \cite{Hoo91}.

Let us briefly explain our notation.
As usual, $\triangle$ and $\cupp$ denote symmetric difference and disjoint union. 
Let $n:=|V|$ and $E={V\choose 2}$; so $(V,E)$ is the complete graph on $V$.
For a vertex set $U\subseteq V$ let $E[U]$ denote the set of edges with both endpoints in $U$,
$\delta(U)$ the set of edges with exactly one endpoint in $U$, 
and $\delta(v):=\delta(\{v\})$ for $v\in V$.
For $x\in\mathbb{R}^E$ and $H\subseteq E$ we write 
$x(H):=\sum_{e\in H} x_e$, $c(x):=\sum_{e=\{v,w\}\in E}c(v,w) x_e$, 
and $c(H):=\sum_{e=\{v,w\}\in H}c(v,w)$.
By $[m]$ we denote the index set $[m]:=\{1,2, \dots, m\}$.
An $A$-$B$-cut is an edge set $\delta(U)$ for some vertex set $U$ with $A\subseteq U\subseteq V\setminus B$.
An \emph{$s$-$t$-tour} is an edge set $H$ such that $(V,H)$ is an $s$-$t$-path (or a circuit if $s=t$).
So if $s\ne t$, $H$ is the edge set of a path with endpoints $s$ and $t$ that spans all vertices.

As all previous works, we use a classical idea of Wolsey \cite{Wol80} for analyzing Christofides' algorithm. 
The following LP is obviously a relaxation of the $s$-$t$-path TSP 
(incidence vectors of $s$-$t$-tours are feasible solutions):
 \begin{equation} \label{eq:subtour_lp_with_degree}
 \begin{aligned}
 & \min c(x) \hspace{-2mm} \\
 &s.t. & x(\delta(U)) &\geq 2  & & \text{for } \emptyset \subset U \subseteq V\setminus\{s,t\},\\
 & & x(\delta(U)) &\geq 1  & & \text{for } \{s\} \subseteq U \subseteq V\setminus\{t\}, \\
 & & x(\delta(v)) &= 2  & & \text{for } v\in V\setminus(\{s\}\triangle\{t\}), \\
 & & x(\delta(v)) &= 1  & & \text{for } v\in \{s\}\triangle\{t\}, \\
 & & x(e) &\geq 0 & & \text{for } e\in E.
\end{aligned}\hspace{-4mm}
\end{equation}
Note that we wrote $\{s\}\triangle \{t\}$ instead of $\{s,t\}$ in order to have a correct formulation even in the case $s=t$;
in this case the second and fourth line of constraints are empty.
Held and Karp \cite{HelK70} observed that every feasible solution to this LP is a 
convex combination of incidence vectors of spanning trees (plus one edge if $s=t$) of $(V,E)$. 
Hence the cost of the cheapest spanning tree $S$ is at most $c(x^*)$ for an optimum LP solution $x^*$.

Our algorithm will not need the degree constraints
and work with the following relaxation:
\begin{equation} \label{eq:subtour_lp}
\begin{aligned}
 & \min c(x) \hspace{-2mm} \\
 &s.t. & x(\delta(U)) &\geq 2  & & \text{for } \emptyset \subset U \subseteq V\setminus\{s,t\},\\
 & & x(\delta(U)) &\geq 1  & & \text{for } \{s\} \subseteq U \subseteq V\setminus\{t\}, \\
 & & x(e) &\geq 0 & & \text{for } e\in E.
\end{aligned}\hspace{-4mm}
\end{equation}
Although we do not need this fact, we remark that both LPs have the same value.\footnote{This 
can be shown with Lovász’ \cite{Lov76} splitting lemma,
as was observed (in a similar context) by Cunningham \cite{MonMP90}, Goemans and Bertsimas \cite{GoeB93}.}

The purpose of the matching $M$ in Christofides' algorithm is to correct the parities of the vertex degrees.
Recall that $T:=\{v\in V:|S\cap \delta(v)| \text{ odd}\}\triangle\{s\}\triangle\{t\}$ is the set of 
vertices with wrong parity.
A $T$-join is an edge set $J$ such that the odd-degree vertices in $(V,J)$ are precisely
the elements of $T$. Since $c$ is a metric, the minimum cost of a matching on $T$ equals the minimum cost of a $T$-join,
and this is the minimum cost of a vector $y$ in the $T$-join polyhedron \cite{EdmJ73}
\begin{equation}\label{eq:Tjoinpolyhedron}
\bigl\{y\in\mathbb{R}^E_{\ge 0} : y(\delta(U))\ge 1 \text{ for $ U\subset V \text{ with } |U\cap T|$ odd} \bigr\}.
\end{equation}

Therefore, the cost of the matching $M$ is at most $c(y)$, for any vector $y$ in this polyhedron.
Since it bounds the cost of parity correction, we call a vector $y$ in \eqref{eq:Tjoinpolyhedron}
a parity correction vector.

If $s=t$, the LP implies $x^*(C)\ge 2$ for every cut $C$ and we can choose $y=\frac{1}{2}x^*$.
Thus we get $c(M) \le \frac{1}{2} c(x^*)$.
This shows an upper bound of $\frac{3}{2}$
on the integrality ratio and on the approximation ratio of Christofides' algorithm \cite{Chr76}. 
This is Wolsey's analysis \cite{Wol80}.

From now on we will assume $s\ne t$.
Call a cut $\delta(U)$ (for $\emptyset\not=U\subset V$) \emph{narrow} if $x(\delta(U))<2$.
Narrow cuts are the reason why Wolsey's argument fails for $s\ne t$.
An, Kleinberg, and Shmoys \cite{AnKS15} showed that the narrow cuts form a chain.
They considered \eqref{eq:subtour_lp_with_degree}, but the degree constraints are not needed and the same proof works:

\begin{proposition}\label{prop:narrow_cuts}
Let $x\in\mathbb{R}_{\ge 0}^E$ be a feasible solution to the linear program \eqref{eq:subtour_lp}.
Then there are $m\ge 0$ sets $X_1,\ldots,X_m$ with 
$\{s\}\subseteq X_1\subset X_2\subset \cdots \subset X_m \subseteq V\setminus\{t\}$ such that 
$$\{\delta(X_i) : i\in[m]\} = \{ \delta(U) : \emptyset\not=U\subset V,\, x(\delta(U))<2 \}.$$
Moreover, all of these sets can be computed by $n^2$ minimum cut computations in the graph $(V,E)$ and 
thus in polynomial time.
\end{proposition}
\begin{proof}
Let $X,Y \subseteq V$ such that $x(\delta(X)) < 2$, $x(\delta(Y)) < 2$ and $s\in X\cap Y$. By the LP constraints 
we have $t\not \in X$ and $t\not \in Y$. Suppose neither $X\subseteq Y$ nor $Y\subseteq X$. Then, $X\setminus Y$ and 
$Y\setminus X$ are both nonempty and contain none of the vertices $s$ and $t$. Thus,
\begin{align*}
 4&> x(\delta(X)) + x(\delta(Y)) \ge x(\delta(X\setminus Y)) + x(\delta(Y\setminus X)) \ge 4,
\end{align*}
a contradiction. To prove that the narrow cuts can be computed efficiently, we observe that for each narrow cut 
$C\in \Nscr$ a pair $\{v,w\}$ of vertices exists such that $C$ is the only narrow cut separating $v$ and $w$.
Thus, by computing a minimum capacity $v$-$w$-cut (with respect to capacities $x$) for all pairs $\{v,w\}$ of vertices we will find all narrow cuts. 
\end{proof}
 
Narrow cuts were the focus of \cite{AnKS15} and all subsequent approximation algorithms (cf.\ Table \ref{table:apxratios}). 
They all also proved upper bounds on the integrality ratio. 
Our recursive dynamic programming approach is completely different.
On a very high level, we guess (by a dynamic program) which of the narrow cuts are crossed only once by an optimum $s$-$t$-tour and which 
are crossed at least three times. 
We partition the instance at the narrow cuts that are crossed only once and strengthen the LP by requiring value at least three 
at the other narrow cuts. We call the dynamic program recursively on the sub-instances.
For any $\epsilon>0$, a fixed number of recursion levels yields the approximation ratio $\frac{3}{2}+\epsilon$.

Very recently, our approach has been improved and simplified by Zenklusen \cite{Zen19}.
He obtained the approximation ratio $\frac{3}{2}$ by considering not only the narrow cuts but all $s$-$t$-cuts 
with value less than three.

Neither our algorithm nor Zenklusen's yields an upper bound on the integrality ratio.
The currently best known upper bound on the integrality ratio is $1.5284$, obtained by a new analysis of the 
Seb\H{o}--van Zuylen algorithm \cite{SebvZ16, TraV18}.

\renewcommand{\arraystretch}{1.3}
\begin{table}
\begin{center}
\begin{tabular}{|l|l|}
\hline
\textbf{reference} & \textbf{ratio} \\
\hline
Hoogeveen \cite{Hoo91} & 1.667\\
\hline
An, Kleinberg, and Shmoys \cite{AnKS15} & 1.618\\
\hline
Seb\H{o} \cite{Seb13} & 1.6\\
\hline
Vygen \cite{Vyg16} & 1.599\\
\hline
Gottschalk and Vygen \cite{GotV16} & 1.566\\
\hline
Seb\H{o} and van Zuylen \cite{SebvZ16} & 1.529\\
\hline
\end{tabular}
\caption{Previous approximation guarantees (rounded).\label{table:apxratios}}
\end{center}
\end{table}

\section{Outline of our algorithm}

We start with a high-level overview, sketching the key idea.

We will compute a spanning tree $(V,S)$ and a parity correction vector in the
$T$-join polyhedron \eqref{eq:Tjoinpolyhedron} for $T:=\{v\in V:|S\cap \delta(v)| \text{ odd}\}\triangle\{s\}\triangle\{t\}$.
The parity correction vector will be a nonnegative combination of LP solutions.
If $x^*_1$ is an optimum solution to the LP \eqref{eq:subtour_lp}, $\frac{1}{2}x^*_1$ would be good,
but it is insufficient for narrow cuts $C$ with $|C\cap S|$ even.
Note that $s$-$t$-cuts $C=\delta(U)$ with $|C\cap S|$ odd are irrelevant because
for these sets $|\{v\in U:|S\cap \delta(v)| \text{ odd}\}|$ is odd and thus
$|U\cap T|= |\{v\in U:|S\cap \delta(v)| \text{ odd}\}\triangle( U \cap \{s,t\})|$ is even.

Let $\Nscr_1$ be the set of narrow cuts of the LP solution $x^*_1$ and let $H$ be a fixed optimum $s$-$t$-tour.
As all narrow cuts are $s$-$t$-cuts, we have for each narrow cut $C$ that $|C\cap H|$ is odd.
Suppose we know the partition $\Nscr_1=\Lscr\cupp\Bscr$ of the narrow cuts into 
\emph{lonely} cuts (cuts $C\in\Nscr_1$ with $|C\cap H|=1$) and
\emph{busy} cuts (cuts $C\in\Nscr_1$ with $|C\cap H|\ge 3$).
Then we can compute a cheapest spanning tree $(V,S)$ with $|S\cap C|=1$ for all lonely cuts $C\in\Lscr$.
This can be easily done in polynomial time because the lonely cuts form a chain.
However, $\frac{1}{2}x^*_1$ is still insufficient for busy cuts.

\renewcommand{\arraystretch}{1.3}
\begin{table}
  \begin{center}
\begin{tabular}{|c|c|c|c|c|c|}
\hline
level & fraction of $x^*_l$  & \multicolumn{4}{c|}{lower bound on LP value} \\
$l$ & in parity & \multicolumn{4}{c|}{$x^*_l(C)$ of busy cuts $C$ for} \\
 & correction vector & $C\in\Nscr_1$ & $C\in\Nscr_2$ & $C\in\Nscr_3$ & $C\in\Nscr_4$ \\
\hline
1 & $\frac{8}{29}$ & 1 & 2 & 2 & 2 \\
2 & $\frac{4}{29}$ & 3 & 1 & 2 & 2 \\
3 & $\frac{2}{29}$ & 3 & 3 & 1 & 2 \\
4 & $\frac{1}{29}$ & 3 & 3 & 3 & 1 \\
\hline
\end{tabular}
\caption{Let $x^*_l$ be the LP solution on level $l$, and $\Nscr_l$ its narrow cuts.
If we enforce $x(C)\ge 3$ for all busy cuts $C\in\Nscr_i$ on all levels $l>i$, 
a nonnegative combination of the LP solutions $x^*_l$ 
with the coefficients in the second column
is a cheap parity correction vector
for any tree $(V,S)$ with $|S\cap C|=1$ for every lonely cut $C$. \label{table:fractions} }
  \end{center}
\end{table} 

Knowing the busy cuts, we can add the constraint $x(C)\ge 3$ for all $C\in\Bscr$ to the LP
and obtain a second solution $x^*_2$. Since $x^*_2(C)$ is big where $x^*_1(C)$ was insufficient,
we can combine the two vectors; for example, $\frac{2}{3}x^*_1+\frac{1}{3}x^*_2$ is an LP solution
with value at least $\frac{5}{3}$ at every cut $C\notin\Lscr$ (while $x^*_1$ could only guarantee $\ge 1$).
The second LP solution $x^*_2$ has new narrow cuts, which again can be lonely or busy.
Adding additional constraints $x(C)\ge 3$ for the new busy cuts, we get a third LP solution $x^*_3$, and so on.
Table \ref{table:fractions} shows how these LP solutions can be combined to a cheap parity correction vector.
(We remark that we could also choose the fractions uniformly, in this case $\frac{1}{7}$ each, but it would require 
more levels to obtain the same approximation ratio.)

  \begin{figure*}
  \begin{center}
   \begin{tikzpicture}[xscale=1.4, yscale=1.2]
    \foreach \x in { 3, 6, 9} {
      \draw [dotted, very thick, Gray] (\x +0.5,0.6) --(\x + 0.5,4);
    }
    \foreach \x in {4, 5} {
      \draw [densely dotted, very thick, Red] (\x +0.5,0.6) --(\x + 0.5,4);
    }
    \foreach \x  in {2, 7, 8,  10} {
      \draw [dashed, ultra thick, darkgreen] (\x +0.5,0.6) --(\x + 0.5,4);
    }  
    \foreach \x  in { 2, 7, 8,  10} {
      \draw [black] (\x +0.55,0.2) --(\x + 0.55,0.4);
      \draw [black] (\x +0.45,0.2) --(\x + 0.45,0.4);
    }
    \draw [black] (1.55,0.2) --(1.55,0.4);
    \draw [black] (11.45,0.2) --(11.45,0.4);
    \foreach \x/\y  in { 1/2, 2/7, 7/8,  8/10, 10/11} {
      \draw[black] (\x + 0.55,0.3) --(\y + 0.45,0.3);
    }    
    \begin{scope}[shift = {(0,0.5)}]
    \foreach \x\y [count=\i] in {4/2, 2/3, 3/1, 4/3, 6/1, 6/3, 7/2, 8/1, 9/3, 10/1, 11/3, 5/1.5, 10/2} {
      \node[circle, fill=none, draw=black, thick, inner sep = 1.2, outer sep = 1.5] (v\i) at (\x,\y) {};
    }
    \foreach \i in {2,4,5,8,9,10,11} {
      \node[circle, fill=black, draw=black, inner sep = 1, outer sep = 1] (v) at (v\i) {};
    }
    \draw[thick, gray] (v4) -- (v6) -- (v1) -- (v3) -- (v12) -- (v7) -- (v5);
    \draw[thick, gray] (v10) -- (v13) -- (v9);
    \begin{scope}[darkgreen, ultra thick] 
      \draw (v2) -- (v4);
      \draw  (v5) -- (v8) -- (v10);
      \draw (v9) -- (v11);
    \end{scope}
    \end{scope}
    \node[left=2pt] at (v2) {$s$};
    \node[right =2pt] at (v11) {$t$}; 
  \end{tikzpicture}
   \caption{The dashed and dotted vertical lines show the narrow cuts. The solid lines show an optimum $s$-$t$-tour.
   The green (bold) edges and the green (dashed) cuts are lonely. The intervals at the bottom indicate the sub-instances 
   of the next recursion level, where the filled vertices serve as $s'$ and/or $t'$.
   (The first, third, and fifth sub-instance consist of a single vertex $s'=t'$.)
   The dotted (red and gray) narrow cuts are busy, but only the red (densely dotted) busy cuts will be passed to the next level
   because they have $s'$ on the left and $t'$ on the right. 
   The gray (loosely dotted) busy cuts will automatically have value at least 3 as the proof will reveal. \label{fig:outline}}
  \end{center}
 \end{figure*}
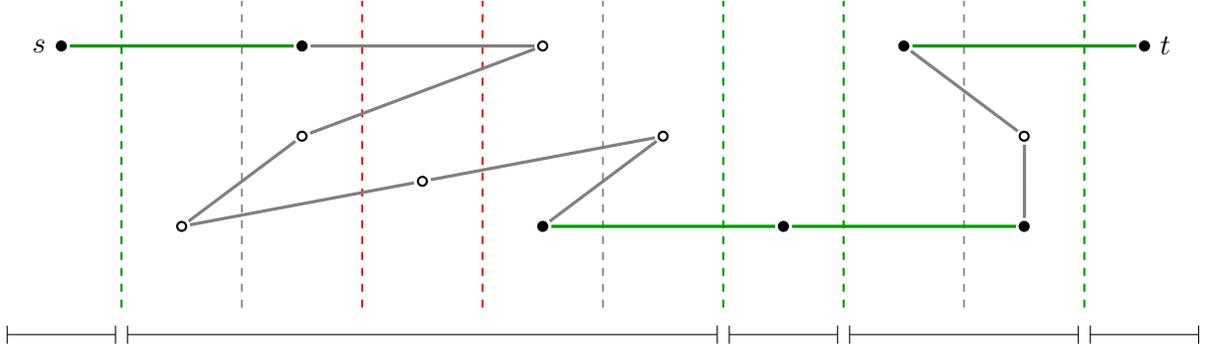
 
If we knew not only the lonely cuts but also the \emph{lonely edges},
i.e. the edge $e\in C\cap H$ for every $C\in\Lscr$, 
then we could partition the original instance at the lonely cuts, solve separate LPs for the sub-instances,
and combine the solutions. See Figure \ref{fig:outline}.

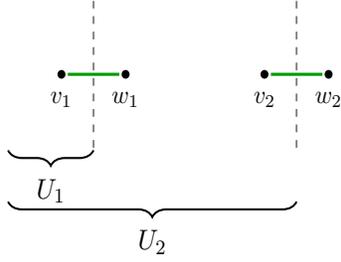
\begin{figure}
   \begin{center}
    \resizebox{0.3\textwidth}{3.5cm}{
    \begin{tikzpicture}[xscale= 1.6, yscale=1]
    \foreach \x in {2.3, 4.2} {
    \draw[ultra thick, dashed, darkgreen] (\x,1) -- (\x,3) ;
    }

    \foreach \x/\n [count=\i from 1] in {2.0/{$v_1$}, 2.6/{$w_1$}, 3.9/{$v_2$}, 4.5/{$w_2$}} {
	\node[circle, fill = black, inner sep = 1.2, outer sep = 1] (v\i) at (\x,2) {};
	\node [below = 3pt] (w) at (v\i) {\small \n};
    }
    \draw[darkgreen, ultra thick] (v1) -- (v2);
    \draw[darkgreen, ultra thick] (v3) -- (v4);

    \draw[thick, decoration={brace,mirror,raise=1pt, amplitude=6pt},decorate] (1.5,1) -- node[below=8pt] {$U_1$} (2.3,1);
    \draw[thick, decoration={brace,mirror,raise=1pt, amplitude=6pt},decorate] (1.5,0.3) -- node[below=8pt] {$U_2$} (4.2,0.3);
    \end{tikzpicture}
    }
    \caption{A possible sub-instance with vertex set $U_2\setminus U_1$, $s'=w_1$, and $t'=v_2$.
    This sub-instance will be represented by the arc $\left((U_1,v_1,w_1), (U_2, v_2, w_2)\right)$ in the digraph $D$.
             Note that the vertices $w_1$ and $v_2$ might be identical. \label{fig:edge_cost}} 
   \end{center}
  \end{figure}

Of course, the main difficulty is that we do not know which cuts are lonely and which are busy, and
we do not know the lonely edges.
However, for each possibility of two subsequent lonely cuts $\delta(U_1)$ and $\delta(U_2)$
with $\{s\}\subseteq U_1\subset U_2 \subseteq V\setminus\{t\}$ and lonely edges $\{v_1,w_1\}$ and $\{v_2,w_2\}$
with $v_1\in U_1$, $w_1,v_2\in U_2\setminus U_1$ and $w_2\in V\setminus U_2$,
we can consider the instance with vertex set $U_2\setminus U_1$ and $s'=w_1$ and $t'=v_2$.
See Figure \ref{fig:edge_cost}.
There are $O(n^4)$ such instances (due to Proposition \ref{prop:narrow_cuts}).
For each such instance we compute a spanning tree and an LP solution (recursively),
and we combine these by dynamic programming.

 The output of the dynamic program is a spanning tree $(V,S)$ and an LP solution $y$. 
 We set $T:= \{ v\in V \mid |\delta(v)\cap S| \text{ odd}\} \triangle \{s\} \triangle\{t\}$, 
 compute a cheapest $T$-join $J$, find an Eulerian trail from $s$ to $t$ in $(V, S \cupp J)$, and shortcut.
 To bound the cost of $J$ we will show that $(\frac{1}{2}+O(2^{-k}))y$ is a parity correction vector,
 where $k$ denotes the number of levels in our recursive dynamic program.
 
 Before we get into the details, let us mention one more subtle point.
 The busy cuts of previous levels can intersect several sub-instances. 
 For a sub-instance on $U_2\setminus U_1$ with $s'=w_1$ and $t'=v_2$, we will only pass a busy cut $C=\delta(U)$ to this sub-instance if 
$U_1\cup\{s'\}\subseteq U \subseteq U_2\setminus\{t'\}$.
For the other busy cuts $C$ (gray in Figure \ref{fig:outline}),
the inequality $x(C)\ge 3$ will follow automatically from combining the LP solutions
returned by the sub-instances and the lonely edges.

 \section{The recursive dynamic program}\label{section:dynamic_program}

 \noindent In this section we describe the dynamic programming algorithm in detail.
  We call the algorithm recursively with a fixed recursion depth $k$. 
  We also fix coefficients $\lambda_1 > \lambda_2 > \dots > \lambda_k > 0$.
  Before describing our algorithm, we first explain the role of these coefficients.
  Our dynamic program yields a spanning tree and an LP solution $y_1$ such that $\lambda_1 y_1$ is 
  a parity correction vector, where $\lambda_1 = \frac{1}{2}+O(2^{-k})$.
  The LP solution $y_1$ will be a convex combination of LP solutions $x^*_1, \dots, x^*_k$ such that 
  \begin{equation}\label{eq:explanation_lambda}
  \lambda_1 \cdot y_1 = (\lambda_1 - \lambda_2) x^*_1 + \dots + (\lambda_{k-1} - \lambda_k) x^*_{k-1} 
     +\lambda_k x^*_k. 
   \end{equation}   
  We will choose the coefficients $\lambda_1, \dots, \lambda_k$ such that 
  $\lambda_1 - \lambda_2 = \frac{2^k}{\Lambda}, \lambda_2 - \lambda_3 = \frac{2^{k-1}}{\Lambda}, \dots, 
  \lambda_{k-1} - \lambda_k = \frac{2}{\Lambda},
  \lambda_k =\frac{1}{\Lambda}$ for some constant $\Lambda > 0$. (See Table \ref{table:fractions} for an example.)
  The coefficient $\lambda_l$ is the total fraction by which the LP solutions $x^*_l, \dots, x^*_k$ contribute 
  to our parity correction vector. More precisely, this contribution is 
  \[ \lambda_l \cdot y_l = (\lambda_l - \lambda_{l+1}) x^*_l +  \dots + (\lambda_{k-1} - \lambda_k) x^*_{k-1} 
     +\lambda_k x^*_k  \]
  for some LP solution $y_l$. 
  In our algorithm we will use the following recursive formula for our parity correction vector~$\lambda_1 y_1$. We have $y_k = x^*_k$ and
  for $l=k-1,\dots, 1$
    \begin{equation*}\label{eq:recursive_formula_y}
      y_l = \frac{\lambda_l - \lambda_{l+1}}{\lambda_l} x^*_l +  \dots + \frac{\lambda_{l+1}}{\lambda_l} y_{l+1}. 
    \end{equation*}
  We give the precise choice of the constants $k$ and $\lambda_i$ ($i\in[k]$)
  depending on $\epsilon$ in Section~\ref{section:analysis_dp}.
  
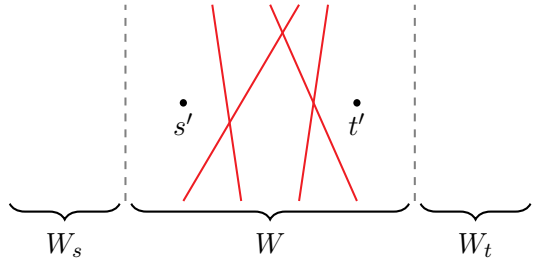
\begin{figure}
   \begin{center}
   \resizebox{0.45\textwidth}{3.5cm}{
     \begin{tikzpicture}[xscale=0.8, yscale=1.3]
    \node[circle, fill = black, inner sep = 1, outer sep = 1] (s') at (3,1) {};
    \node[circle, fill = black, inner sep =1, outer sep = 1] (t') at (6,1) {};

    \node [below] (v) at (s') {$s'$};
    \node [below] (v) at (t') {$t'$};

    \draw[ultra thick, dashed, darkgreen] (2,0) -- (2,2);
    \draw[ultra thick, dashed, darkgreen] (7,0) -- (7,2);

    \draw[thick,Red] (3,0) -- (5,2);
    \draw[thick,Red] (4,0) -- (3.5,2);
    \draw[thick,Red] (6,0) -- (4.5,2);
    \draw[thick,Red] (5,0) -- (5.5,2);

  \draw[thick, decoration={brace,mirror,raise=1pt, amplitude=6pt},decorate] (0,0) -- node[below=8pt] {$W_s$} (1.9,0);
  \draw[thick, decoration={brace,mirror,raise=1pt, amplitude=6pt},decorate] (2.1,0) -- node[below=8pt] {$W$} (6.9,0);
  \draw[thick, decoration={brace,mirror,raise=1pt, amplitude=6pt},decorate] (7.1,0) -- node[below=8pt] {$W_t$} (9,0);
 \end{tikzpicture}
 }
   \end{center}
   \caption{The input to the dynamic program.
    The dashed lines are the cuts $\delta(W_s)$, and  $\delta(W_t)$.
    The solid red lines indicate possible busy cuts, i.e. elements of $\Bscr$.
   \label{fig:input_dp}}
  \end{figure}

  Now we describe the dynamic programming algorithm.
  The input to the dynamic program (see Figure \ref{fig:input_dp}) consists of
  \begin{itemize}
  \item sets $W_s,W_t \subseteq V$ with $W_s \cap W_t = \emptyset$;
  \item vertices $s',t'\in W:=V\setminus (W_s \cup W_t)$; note that $s'=t'$ is possible;
  \item a collection $\Bscr$ of busy $(W_s \cup \{s'\})$-$(W_t\cup \{t'\})$-cuts; and
  \item a level $l\in[k]$.
  \end{itemize}
  
  The output of the dynamic program is 
  \begin{itemize}
   \item a tree $(W,S)$;
   \item a vector $y\in \mathbb{R}^E_{\ge 0}$, which will contribute to the parity correction vector; and
   \item a chain $\Lscr$ of $(W_s \cup \{s'\})$-$(W_t\cup \{t'\})$-cuts with $|S\cap C|=1$ for all $C\in \Lscr$.
  \end{itemize}
  We remark that for computing an $s$-$t$-tour it is sufficient to return the tree $(W,S)$ and the cost of the vector $y$. 
  The chain $\Lscr$ and the explicit vector $y$ are added only for the purpose of analysis.
  
  The dynamic programming algorithm first computes an optimum solution $x^*$ to the following linear program.
  
   \begin{equation} \label{eq:dynamic_program_LP}
   \begin{aligned}
   & \min c(x) \hspace{-2mm} \\
   & s.t. & x(\delta(U)) &\geq 2  & & \text{for }\emptyset \not =U\subseteq W\setminus\{s',t'\} \\
   & & x(\delta(U)) &\geq 1  & & \text{for } \{s'\}\subseteq U \subseteq W\setminus\{t'\} \\
   & &  x(C) &\ge 3 \ \ & & \text{for }  C\in \Bscr \\
   & &  x(e) &\geq 0 & & \text{for } e\in E[W] \\
   & &  x(e) &= 0  & & \text{for } e\in E\setminus E[W]. 
   \end{aligned}\hspace{-6mm}
   \end{equation}
  The vector $x^*$ restricted to edges $e\in E[W]$ is a feasible solution 
  of linear program \eqref{eq:subtour_lp} for the instance of the 
  metric $s$-$t$-path TSP with vertex set $W$ and $s=s'$ and $t=t'$. 
  It is still useful that $x^*$ is a vector in the entire space $\mathbb{R}^E$ because we will add vectors for 
  different sub-instances later.
  
  We consider the relevant set of \emph{narrow cuts} 
  \begin{align*}
  \Nscr:= \big\{ \delta(U) \mid\ & x^*(\delta(U))< 2,  
   W_s \cup \{s'\} \subseteq U \subseteq V\setminus \left(W_t \cup \{t'\}\right)\big\}.
  \end{align*}
  By Proposition \ref{prop:narrow_cuts}, $\Nscr$ forms a chain, i.e., there exist sets 
  \[ W_s \cup  \{ s'\} \subseteq X_1 \subset X_2 \subset \dots \subset X_m \subseteq 
  V \setminus \left(W_t \cup \{t'\}\right)\]
  such that $\Nscr = \{ \delta (X_i) \mid i \in [m]\}$. 
  
  If we have $l=k$, i.e. we are on the final level $k$, we return the vector $y:= x^*$ 
  and a minimum cost tree $(W,S)$.
  Moreover, we return $\Lscr = \emptyset$.
  
  Otherwise, i.e. if $l<k$, we will apply our algorithm recursively to all possible sub-instances that could 
  occur by partitioning $\Nscr$ into busy and lonely cuts and choosing lonely edges.
  Then we combine these sub-instances optimally.
  There can be exponentially many ways to combine the sub-instances, but we can find an optimum combination
  by dynamic programming. We describe the dynamic program as a shortest path search in a directed auxiliary graph $D$.
  The vertices of $D$ correspond to the different states/table entries of the dynamic program and the 
  arcs correspond to possible sub-instances.  
  More precisely, we construct a directed auxiliary graph $D$ with vertices
  \begin{align*}
     V(D) := \big\{& (U,v,w) \mid \delta(U) \in \Nscr, s'\in U, v\in U\cap W, 
              w\in W\setminus U  \big\}
              \cupp \big\{ (W_s, \emptyset, s'), (V\setminus W_t, t', \emptyset) \big\} \\
   \intertext{and arcs }
   E(D) := &\bigl\{\left((U_1,v_1,w_1), (U_2, v_2, w_2)\right) \mid 
     U_1 \subset U_2, w_1,v_2\in U_2 \setminus U_1 \bigr\}. 
  \end{align*}
  Figure \ref{fig:edge_cost} illustrates the sets and vertices defining an arc $a\in E(D)$.
  Every $(W_s,\emptyset,s')$-$(V\setminus W_t,t',\emptyset)$-path in the auxiliary digraph $D$
  corresponds to a possible combination of sub-instances (corresponding to the arcs of the path) or, equivalently, 
  to a possible guess of lonely cuts and lonely edges (corresponding to the inner vertices of the path).

    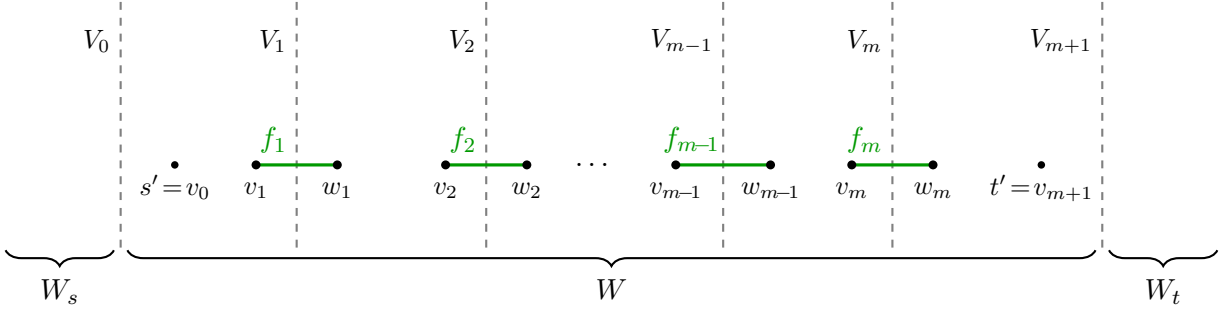
\begin{figure*}
   \begin{center}
    \begin{tikzpicture}[xscale= 1.6, yscale=1.1]
      \foreach \x/\i in {1.5/0,8.75/{m+1}} {
      \draw[ultra thick, dashed, darkgreen] (\x,0) -- (\x,3) ;
      \node[left] (v\i) at (\x,2.5) {\small $V_{\i}$};
      }
       \foreach \x/\i in {2.8/1, 4.2/2, 5.95/{m-1}, 7.2/{m}} {
      \draw[thick, dashed, darkgreen] (\x,0) -- (\x,3) ;
      \node[left] (v\i) at (\x,2.5) {\small $V_{\i}$};
      }

      \node[circle, fill = black, inner sep = 1, outer sep = 1] (s') at (1.9,1) {};
      \node[circle, fill = black, inner sep =1, outer sep = 1] (t') at (8.3,1) {};

      \node [below] (v) at (s') {\small$s'\!=\!v_0$};
      \node [below] (v) at (t') {\small$t'\!=\!v_{m+1}$};

      \node () at (5,1) {$\dots$};

      \draw[ultra thick, darkgreen] (2.5,1)-- (3.1,1);
      \draw[ultra thick, darkgreen] (3.9,1)-- (4.5,1);
      \draw[ultra thick, darkgreen] (5.6,1)-- (6.3,1);
      \draw[ultra thick, darkgreen] (6.9,1)-- (7.5,1);

      \node[darkgreen, left] (v) at (2.8,1.3){$f_1$};
      \node[darkgreen, left] (v) at (4.2,1.3){$f_2$};
      \node[darkgreen, left] (v) at (6.0,1.3){$f_{m\!-\!1}$};
      \node[darkgreen, left] (v) at (7.2,1.3){$f_m$};
      \foreach \x/\n in {2.5/{$v_1$}, 3.1/{$w_1$}, 3.9/{$v_2$}, 4.5/{$w_2$}, 5.6/{$v_{m\!-\!1}$}, 6.3/{$w_{m\!-\!1}$},
	  6.9/{$v_m$}, 7.5/{$w_m$}} {
	  \node[circle, fill = black, inner sep = 1.2, outer sep = 1] (v) at (\x,1) {};
	  \node [below = 3pt] (w) at (v) {\small \n};
      }

      \draw[thick, decoration={brace,mirror,raise=1pt, amplitude=6pt},decorate] (0.65,0) -- node[below=8pt] {$W_s$} (1.45,0);
      \draw[thick, decoration={brace,mirror,raise=1pt, amplitude=6pt},decorate] (1.55,0) -- node[below=8pt] {$W$} (8.7,0);
      \draw[thick, decoration={brace,mirror,raise=1pt, amplitude=6pt},decorate] (8.8,0) -- node[below=8pt] {$W_t$} (9.6,0);
     \end{tikzpicture}
   \caption{The dashed lines show the cuts $\delta(V_j)$ for $j=0,1,\dots,m+1$, where the sets $V_j$ are 
   the sets left of the dashed lines. 
   The partition of the vertex set into $W_s$, $W_t$ and $W$ is shown at the bottom of the picture. 
   The edges $f_j$ are drawn in green. 
   We remark that the vertices $w_j$ and $v_{j+1}$ might be identical for $j=0, 1, \dots , m$. 
   \label{fig:definition_V_j_f_j}}
   \end{center}
  \end{figure*}   
  
  The next step of the algorithm is to compute weights for the arcs of the digraph $D$. 
  For an arc \[a=\left((U_1,v_1,w_1), (U_2, v_2,w_2)\right) \in E(D)\] we define 
  \[ \Bscr^{a} := \left\{\phantom{\bigm.}\delta(U)\in \Nscr \cup \Bscr \mid U_1\cup\{w_1\} 
  \subseteq U \subseteq U_2\setminus\{v_2\} \phantom{\bigm.}\right\}.\] 
  We call the dynamic program  with
  \begin{itemize}
   \item $W_s = U_1$ and $W_t= V\setminus U_2$,
   \item $s' =w_1$ and $t' =v_2$, 
   \item $\Bscr$ = $\Bscr^{a}$, and
   \item the level $l+1$.
  \end{itemize}
  Let the output of this application of the dynamic program be the tree $(U_2\setminus U_1, S^a)$, 
  the vector $y^a \in \mathbb{R}_{\ge 0}^E$, 
  and the chain $\Lscr^a $ of cuts $C$.
  Then we define the cost of the arc $a\in E(D)$ to be
  \begin{equation} \label{eq:def_edge_cost}
   d(a):= \begin{cases}
           c\left(S^a\right) + \lambda_{l+1} \cdot c\left(y^a\right) +\left(1+\lambda_{l+1} \right)\cdot c(v_2,w_2),& 
           \text{if }w_2 \not = \emptyset\\[2mm]
           c\left(S^a\right) + \lambda_{l+1} \cdot c\left(y^a\right),& \text{if } w_2 = \emptyset. 
          \end{cases}
  \end{equation}
  Before we explain the reason for choosing the arc costs like this, we complete the description of our algorithm.
  
  We compute a shortest $(W_s,\emptyset,s')$-$(V\setminus W_t,t',\emptyset)$-path $P$ in the auxiliary digraph $D$
  with respect to the arc costs $d$.
  Let $(W_s, \emptyset,s')=(V_0, v_0,w_0)$, $(V_1, v_1,w_1)$, $(V_2,v_2,w_2)$, $\dots$, 
  $(V_m, v_m,w_m)$, $(V_{m+1}, v_{m+1},w_{m+1}) = (V\setminus W_t,t',\emptyset)$ 
  be the vertices of the path $P$ visited in exactly this order (see Figure~\ref{fig:definition_V_j_f_j}).
  We denote the arcs of $P$ by \[ a_j:=((V_j,v_j,w_j), (V_{j+1}, v_{j+1},w_{j+1}))\ (j=0,\ldots,m).\]
  Moreover, for every $j\in[m]$ let $f_j:=\{v_j,w_j\}$. 
    
  We combine the spanning trees of the sub-instances and the guessed lonely edges to a spanning tree $S$ on the entire set $W$:
  \begin{align*}
   S:= &\left\{ e\in S^a \mid a\in E(P) \right\}   \cup \left\{f_j : j\in [m]\right\}.
  \end{align*}
  Similarly we combine the LP solutions: let 
  \begin{equation*}
   y':= \sum_{a\in E(P)} y^a + \sum_{j\in [m]} \chi^{f_j},
  \end{equation*}
  where $\chi^{f_j}$ is the incidence vector of $f_j$ (i.e., $\chi^{f_j}_{f_j}=1$ and $\chi^{f_j}_e = 0$ for $e\in E\setminus\{f_j\}$).
  
  Define $y$ to be the following convex combination of $x^*$ and $y'$:
  \begin{equation*}
   y:= \frac{\lambda_l - \lambda_{l+1}}{\lambda_l} \cdot x^* + \frac{\lambda_{l+1}}{\lambda_l} \cdot y'.
  \end{equation*}
  We set 
  \begin{align*}
   \Lscr :=\{ C \mid C \in \Lscr^a \text{ for some }a\in E(P) \} \cup \{\delta(V_j) \mid j\in[m]\}
  \end{align*}
  and return the edge set $S$, the vector $y$ and the set $\Lscr $.
  
  We can now give intuition for the arc costs $d$. 
  The contribution of arc $a=\left((U_1,v_1,w_1), (U_2, v_2,w_2)\right)$ of $P$ 
  to the spanning tree $S$ consists of $S^a$ and the edge $\{v_2,w_2\}$ (if $w_2 \ne \emptyset$).
  The contribution to the parity correction vector is $\lambda_{l+1}(y^a +  \chi^{\{v_2,w_2\}})$
  because $\lambda_{l+1}$ is the total fraction by which the LP solutions of levels $l+1,\dots,k$ 
  contribute to our parity correction vector. See \eqref{eq:explanation_lambda}.
  
  \section{Properties of the dynamic program}\label{section:analysis_dp}
  
  In this section we show several important properties of the output of the dynamic program.
  We show all these properties by induction on $k-l$, i.e. 
  to prove them we assume that they hold for all levels $l'$ with $l<l'\le k$.
  First, we show that the set $\Lscr$ of all guessed lonely cuts (in all levels) forms a chain.
      
  \begin{lemma}\label{lemma:Lscr_chain}
   $\Lscr$ is a chain of $(W_s \cup \{s'\})$-$(W_t \cup \{t'\})$-cuts.
  \end{lemma}
  \begin{proof} If $l=k$, we have $\Lscr = \emptyset$. So we may assume $l<k$.
  If a cut $C$ belongs to $\Lscr$, it is a cut $\delta(V_j)$ for some $j\in[m]$ or is contained in $\Lscr^a$ 
  for some $a\in E(P)$. Recall that
  \[ W_s = V_0 \subset V_1 \subset V_2 \subset \dots \subset V_m \subset V_{m+1} = V \setminus W_t. \]
  Moreover, all cuts $\delta(V_j)$ for $j\in [m]$ are in the set $\Nscr$ of narrow cuts, which implies 
  \[W_s \cup \{s' \} \subseteq V_j \subseteq  V\setminus \left(W_t \cup \{t'\}\right).\]
  
  Now consider the cuts $\Lscr^{a_j}$ for $j\in\{0,1,\dots,m\}$.
  By induction on $k-l$, the cuts in $\Lscr^{a_j}$ are a chain of cuts of the form $\delta(U)$ 
  for a set $U$ with $V_j \cup \{w_j\} \subset U \subset V_{j+1} \setminus \{v_{j+1}\}$.
  Since $s' \in V_j \cup \{w_j\}$ and $t' \notin V_{j+1} \setminus \{v_{j+1}\}$,
  these cuts are $(W_s \cup \{s'\})$-$(W_t \cup \{t'\})$-cuts.
  Moreover,  $\{\delta(V_j) \mid j\in[m] \} \subseteq \Nscr$ 
  remains a chain when adding the sets $\Lscr^a$ for all $a\in E(P)$.
  \end{proof}
  
  Next, we show that each of our guessed lonely edges belongs to only one guessed lonely cut.
  \begin{lemma}\label{lemma:f_j_not_in_Lscr_a}
   For $l<k$, an edge $f_j$ with $j\in [m]$ is not contained in any cut $C\in \Lscr^a$ for $a\in E(P)$.
  \end{lemma}
  \begin{proof} 
  Assume an edge $f_j$ for $j\in[m]$ is contained in a cut $C\in\Lscr^a$ for some $a\in E(P)$. 
  As the edge $f_j$ is contained in neither $\delta(V_{j-1})$ nor $\delta(V_{j+1})$,
  one endpoint is in $V_j \setminus V_{j-1}$ and the other endpoint is in $V_{j+1} \setminus V_j$. 
  Using Lemma \ref{lemma:Lscr_chain}, this implies $a=a_{j-1}$ or $a=a_j$.
  If $a=a_{j-1}$, the endpoint $v_j$ of $f_j$ is contained in $V_j$ and 
  plays the role of $t'$ in the dynamic program computing the tree $S^a$. 
  This  implies by Lemma \ref{lemma:Lscr_chain} that for a cut $C\in \Lscr^a$ we have $C=\delta(U)$ for some $U$ with
  $V_{j-1} \subseteq U \subseteq V_j \setminus \{v_j\}$, and hence $f_j \not \in C=\delta(U)$.
  For the case $a=a_j$ a symmetric argument shows $f_j \not \in C$ for $C\in \Lscr^{a_j}$.
  \end{proof}
  
  Now we show that we indeed construct a spanning tree that crosses the guessed lonely cuts only once.
  \begin{lemma}\label{lemma:T_cuts}
  The graph $(W,S)$ is a tree.
  For every cut $C\in \Lscr$ we have $|S\cap C| = 1$.
  \end{lemma}
  \begin{proof}
  For level $l=k$ the chain $\Lscr$ is empty, and hence the statement is trivial. So assume $l < k$. 
  
  By the construction of the digraph $D$ we have 
  $W_s = V_0 \subset V_1 \subset V_2 \subset \dots \subset V_m \subset V_{m+1} = V \setminus W_t$.
  We have $f_j \in \delta(V_j)$ and $f_j \not\in \delta(V_h)$ for $h\not = j$.
  By induction, $(V_{j+1} \setminus V_j, S^{a_j})$ is a tree for every $j \in \{0,1,\dots,m\}$.
  The edges $f_j$ (for $j\in [m]$) connect these trees to a tree spanning $W$. 
  We observe that $S\cap \delta(V_j) = \{ f_j \}$ for every $V_j $ with $j\in [m]$.
  
  By induction we have $|S^a \cap C|=1$ for all $a\in E(P)$ and $C\in \Lscr^a$. 
  Moreover, note that edges of $S^a$ are not contained in any cut $C\in \Lscr \setminus \Lscr^a$.
  As observed above, the tree $(W,S)$ is constructed such that 
  $S\cap \delta(V_j) = \{ f_j \}$ for every $j\in [m]$.
  Thus, it only remains to show that an edge $f_j$ for $j\in[m]$ 
  can not be contained in a cut $C\in\Lscr^a$ for any $a\in E(P)$ 
  which is precisely the statement of Lemma \ref{lemma:f_j_not_in_Lscr_a}.
  \end{proof}
  \bigskip
  
  Now we bound the cost of the spanning tree $S$ and the contribution $\lambda_l \cdot y$ to the parity correction vector.
  \begin{lemma}\label{lemma:path_length_equals_cost}
   For levels $l<k$ the cost $d(P)$ of the path $P$ equals
   the cost $c(S) + \lambda_{l+1} \cdot c(y')$ of the tree $S$ and the vector $\lambda_{l+1} \cdot y'$.
  \end{lemma}
  \begin{proof} 
  We have 
  \begin{align*}
   c(S) &= \sum_{a\in E(P)}  c(S^a) + \sum_{j=1}^m c(f_j),
   \intertext{and}
   \lambda_{l+1} \cdot c(y') 
   &= \lambda_{l+1} \cdot\sum_{a\in E(P)} c\left(y^a\right) + \lambda_{l+1} \cdot \sum_{j=1}^m c(f_j).
  \end{align*}
  Together with the definition \eqref{eq:def_edge_cost} of the arc cost in $D$ this shows  
  $d(P) = c(S) + \lambda_{l+1} \cdot c(y')$.
  \end{proof}
  
  We fix an optimum $s$-$t$-tour $H$.
  We say an input  $W_s, W_t, s', t', \Bscr$ to the dynamic program is \emph{consistent} with the tour $H$
  if $H$ (traversed from $s$ to $t$) visits $s'$ before $t'$ and the $s'$-$t'$-path in $H$ contains
  exactly the vertices in $V\setminus(W_s\cup W_t)$  and $|H\cap C| \not = 1$ for every cut $C\in \Bscr$.
  We say that a path $\bar P$ in the auxiliary digraph $D$ is \emph{consistent} with the tour $H$ if 
  \begin{itemize}
   \item $\delta(U) \cap H = \{\{v,w\}\}$ for every $(U,v,w)\in V_{\text{in}}(\bar P)$, and
   \item for every cut $C\in \Nscr \setminus \{\delta(U) \mid (U,v,w) \in V_{\text{in}}(\bar P)\}$ we have $|H\cap C| \not = 1$,
  \end{itemize}
  where $V_{\text{in}}(\bar P)$ denotes the set of inner vertices of the path $\bar P$.
  Note that for parity reasons $|H\cap C| \not = 1$ implies $|H\cap C| \ge 3$ for every $s$-$t$-cut $C$.

  We denote by $H_{[s',t']}$ the edge set of the unique path from $s'$ to $t'$ that is contained in the path $(V,H)$.

  \begin{lemma}\label{lemma:bound_on_cost}
   If the input to the dynamic program is consistent with the tour $H$, we have 
   \[ c(S) + \lambda_l \cdot c(y) \le (1+\lambda_l) \cdot c\left(H_{[s',t']}\right). \]
  \end{lemma}
  
  \begin{proof} 
  If the input of the dynamic program is consistent with the tour $H$, the incidence vector of $H_{[s',t']}$ 
  is a feasible solution to the linear program \eqref{eq:dynamic_program_LP} and thus
  \begin{equation}\label{eq:x*_bounded_by_opt}
   c(x^*) \le c\left(H_{[s',t']} \right).
  \end{equation}
 
 For $l=k$ we therefore have $c(y) =c(x^*) \le c(H_{[s',t']})$; 
 moreover  $(W,H_{[s',t']})$ is a tree and therefore we have $c(S)\le c(H_{[s',t']})$.
  
  Now assume $l<k$. 
  Let $\bar P$ be the unique $(W_s,\emptyset,s')$-$(V\setminus W_t,t',\emptyset)$-path in $D$ whose set of inner vertices 
  is exactly the set of vertices $(U,v,w)\in V(D)$ with $\{\{v,w\}\}= H\cap \delta(U)$.
  Then $\bar P$ is consistent with the tour $H$.
 
  For $a=((U_1,v_1,w_1),(U_2,v_2,w_2))\in E(\bar P)$ let $s^a:=w_1$ and $t^a:=v_2$.
  The tour $H$ is the disjoint union of the $H_{[s^a, t^a]}$ 
  for $a\in E(\bar P)$ and the edges $\{v,w\}$ for $(U,v,w)\in V_{\text{in}}(\bar P)$.
  By induction on $k-l$, we have 
  \begin{equation*}
    c\left(S^{a}\right) + \lambda_{l+1} \cdot c\left(y^{a}\right) \le \left(1+\lambda_{l+1}\right) 
    \cdot c\left(H_{[s^a, w^a]} \right).                
  \end{equation*}
  Hence, 
  \begin{equation*}
   \begin{aligned}
    d(\bar P) =\ &\sum_{a\in E(\bar P)} c\left(S^{a}\right) 
                    + \lambda_{l+1}  \sum_{a\in E(\bar P)} c\left(y^{a}\right)  
    + (1+ \lambda_{l+1}) \sum_{(U,v,w)\in V_{\text{in}}(\bar P)} c(v,w) 
                    \\[2mm]
                \le\  &\sum_{a\in E(\bar P)} \left(1+\lambda_{l+1}\right) \cdot c\left(H_{[s^a, t^a]} \right) 
                       + \sum_{(U,v,w)\in V_{\text{in}}(\bar P)} \left(1 + \lambda_{l+1}\right) \cdot c(v,w) \\[2mm]
                =\ &\left(1 + \lambda_{l+1}\right) \cdot c\left(H_{[s',t']}\right).
   \end{aligned}
  \end{equation*}
  Using Lemma \ref{lemma:path_length_equals_cost} and the fact that $P$ is no longer than $\bar P$ we get
  \begin{align*}
   c(S)+\lambda_{l+1} \cdot c(y') &= d(P) 
   \le d(\bar P) 
   \le \left(1 + \lambda_{l+1}\right) \cdot c\left(H_{[s',t']}\right).
  \end{align*}
  
  Using also \eqref{eq:x*_bounded_by_opt} and
  \[ \lambda_l \cdot y = \lambda_{l+1} \cdot y' + \left(\lambda_l - \lambda_{l+1}\right) \cdot x^*   \]
  we get 
  \begin{align*}
     c(S) + \lambda_l \cdot c(y) =\ & c(S) + \lambda_{l+1} \cdot c(y') 
     + \left(\lambda_l - \lambda_{l+1}\right) \cdot c(x^*)  \\[2mm]
   \le\ &
   \left(1 + \lambda_{l+1}\right) \cdot c\left(H_{[s',t']}\right) 
   + \left(\lambda_l - \lambda_{l+1}\right) \cdot c\left(H_{[s',t']} \right) 
   \\[2mm]
   =\ & (1+\lambda_l) \cdot c\left(H_{[s',t']}\right).\\[-10mm]
  \end{align*}
  \end{proof}
  
  The remaining lemmas of this section will be needed to prove that we obtain a 
  feasible parity correction vector. 
  \begin{lemma}\label{lemma:y_1_for_V_j_cuts}
   If $l<k$, the support of the vector $y'$ is a subset of $E[W]$ and we have
   $y'(\delta(V_j)) =1$  for every cut $\delta(V_j)$ with $j\in [m]$.
  \end{lemma}
  \begin{proof}
  The vector $y'$ is defined as the sum of vectors with support contained in $E[W]$.
  Thus, also the support of $y'$ is a subset of $E[W]$.  
  Next, we prove $y'(\delta(V_j)) = 1$ for every cut $\delta(V_j)$ with $j\in [m]$.
  We have $E(P) =\{a_j \mid j\in \{0,1,\dots,m\}\}$ and for every edge $a_j$ the support of $y^{a_j}$ is contained in 
  $E[V_{j+1}\setminus V_j]$. 
  Thus, for every pair of indices $j,r \in \{0,1,\dots,m\}$ we have $y^{a_j}(\delta(V_r)) = 0$.
  As an edge $f_r$ is contained in $\delta(V_r)$, but not in any other cut $\delta(V_j)$ with $j\not = r$,
  we have $y'(\delta(V_j)) = y'(f_j) = 1$. 
  \end{proof}
  
  \begin{lemma}\label{lemma:y_lp_solution}
   The vector $y'$ (for $l<k$) and the vector $y$ are feasible solutions to the following linear program:
  \begin{equation}\label{eq:lp_from_lemma_dp}
   \begin{aligned}
   & \min c(x) \hspace{-2mm} \\
   & s.t. & x(\delta(U)) &\geq 2  & & \text{for }\emptyset \not =U\subseteq W\setminus\{s',t'\} \\
   & & x(\delta(U)) &\geq 1  & & \text{for } \{s'\}\subseteq U \subseteq W\setminus\{t'\} \\
   & & x(e) &\geq 0 & & \text{for } e\in E[W] \\
   & & x(e) &= 0  & & \text{for } e\in E\setminus E[W].
   \end{aligned}\hspace{-4mm}
   \end{equation}
  \end{lemma}
  \begin{proof}
  The vector $x^*$ is a feasible solution to the linear program \eqref{eq:dynamic_program_LP}, and hence,
  also a solution to \eqref{eq:lp_from_lemma_dp}. If $l=k$, we have $y=x^*$, completing the proof for this case.
  We now assume $l<k$ and show, that also $y'$ is a solution to \eqref{eq:lp_from_lemma_dp}. 
  As $y$ is a convex combination of $x^*$ and $y'$, this implies the statement of the Lemma.
  
  The vector $y'$ is defined as the sum of nonnegative vectors with support contained in $E[W]$,
  so $y'\ge 0$ and $y'(e)=0$ for $e\in E\setminus E[W]$.
  It remains to check the cut constraints.
      \begin{figure}
   \begin{center}
   \begin{tikzpicture}[yscale=0.7, xscale=0.65]
    \begin{scope}
      \clip (1.5,0) rectangle (4,4);
      \fill[SkyBlue, opacity=0.4] (6.5,2) ellipse ({4.3} and {1.5});
    \end{scope}
    \begin{scope}
      \clip (9,0) rectangle (11.5,4);
      \fill[SkyBlue, opacity=0.4] (6.5,2) ellipse ({4.3} and {1.5});
    \end{scope}
    \foreach \x/\j in { 1.5/{j_{\min}}, 4/{j_{\min}+1}, 9/{j_{\max}}, 11.5/{j_{\max}+1} } {
      \draw[thick, dashed, darkgreen] (\x,0) -- (\x,4) ;
      \node[below] at (\x, -0.1) {\scriptsize$\delta(V_{\j})$};
      }
    \draw[Blue, thick] (6.5,2) ellipse ({4.3} and {1.5});
    \node[Blue] at (5,3.7) {$U$};

    \end{tikzpicture}
    \caption{The picture illustrates the definition of $j_{\min}$ and $j_{\max}$. 
    The dashed lines show the cuts written below. 
    The indices $j_{\min}$ and $j_{\max}$ are chosen such that the two light blue subsets are both
    nonempty. \label{fig:definition_j_min_max}}
   \end{center}
  \end{figure}
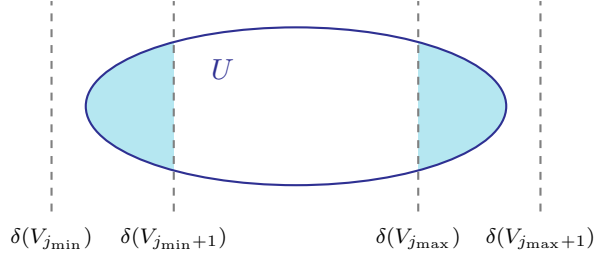
  
  First consider $\delta(U)$ with $\{s'\}\subseteq U\subseteq W\setminus\{t'\}$.
  If there exists an index $j\in \{0,1,\dots,m\}$ such that $(V_{j+1}\setminus V_j) \cap U$ and 
  $(V_{j+1}\setminus V_j) \setminus U$  are both not empty, we have 
  $y'(\delta(U)) \ge y^{a_j}(\delta(U)) \ge 1$ by induction.
  Otherwise, there exists an index $j\in[m]$ such that $\delta(U)$ separates the sets $V_{j+1}\setminus V_j$ 
  and $V_{j}\setminus V_{j-1}$.
  Then, the edge $f_j$ is contained in $\delta(U)$, implying $y'(\delta(U)) \ge \chi^{f_j}(C) \ge 1$.

  Now consider $\delta(U)$ with $\emptyset\not= U\subseteq W\setminus\{s',t'\}$.
  
  Let $j_{\min} \in \{0,1,\dots m\}$ be the minimal index such that $(V_{j_{\min}+1}\setminus V_{j_{\min}}) \cap U$ is nonempty and 
  $j_{\max} \in \{0,1,\dots m\}$ the maximal index such that $(V_{j_{\max}+1}\setminus V_{j_{\max}}) \cap U$ is nonempty
  (see Figure \ref{fig:definition_j_min_max}).

   \begin{figure*}
   \begin{center}
   \begin{tikzpicture}[yscale=0.9, xscale=0.75]
    \foreach \s/\c/\q in {10/9.7/11.5, 14.8/14.6/16.4, 6/4.4/6.1, 1/-0.8/1.0} {
    \begin{scope}
      \clip (\c,0) rectangle (\q,4);
      \fill[SkyBlue, opacity=0.4] (\s - 0.1,2) ellipse ({1.2} and {0.9});
    \end{scope}
    }

    \foreach \s in {10,14.8,6,1} {
      \draw[Blue, thick] (\s - 0.1,2) ellipse ({1.2} and {0.9});
      \node[Blue] at (1.1+\s,1.1) {\small$U$};
    }
    \foreach \s/\l in {0/a,4.8/b,9.7/c,14.6/d} {
      \node[black] at (\s-1.5,3.2) {(\l)};
    }

    \node[darkgreen] at (3.8,1.7) {\scriptsize$f_{j_{\min}}$};
    \node[darkgreen] at (12.2,1.7) {\scriptsize$f_{j_{\max}+1}$};

    \foreach \x/\j in {9.7/{j_{\max}}, 11.5/{j_{\max}+1}, 14.6/{j_{\max}}, 16.4/{j_{\max}+1}, 
		      4.4/{j_{\min}}, 6.1/{j_{\min}+1}, -0.8/{j_{\min}}, 1.0/{j_{\min}+1}} {
      \draw[thick, dashed, darkgreen] (\x,0.5) -- (\x,3.5) ;
      \node[below] at (\x, 0.4) {\scriptsize$\delta(V_{\j})$};
      }

    \draw[very thick, darkgreen] (10.4,2.2)-- (11.9,2.2);
    \draw[very thick, darkgreen] (4.0,2.2)-- (5.5,2.2);

    \node[circle, fill = black, inner sep = 1.2, outer sep = 1] (v) at (10.4,2.2) {};
    \node [below = 3pt] at (v) {\scriptsize $v_{j_{\max}+1}$};
    \node[circle, fill = black, inner sep = 1.2, outer sep = 1] (v) at (15.7,2.8) {};
    \node [above = 2pt] at (v) {\scriptsize $v_{j_{\max}+1}$};
    \node[circle, fill = black, inner sep = 1.2, outer sep = 1] (v) at (5.5,2.2) {};
    \node [below = 3pt] at (v) {\scriptsize $w_{j_{\min}}$};
    \node[circle, fill = black, inner sep = 1.2, outer sep = 1] (v) at (-0.2,2.8) {};
    \node [above = 2pt] at (v) {\scriptsize $w_{j_{\min}}$};

      \foreach \x in {11.9,4.0} {
	\node[circle, fill = black, inner sep = 1.2, outer sep = 1] (v) at (\x,2.2) {};
    }

    \end{tikzpicture}
    \caption{Different cases occurring in the proof of Lemma \ref{lemma:y_lp_solution}.
    The dashed vertical lines indicate the cuts written below. The set $U$ is shown in blue. 
    The light blue subset is nonempty. \label{fig:cases}}
   \end{center}
  \end{figure*}
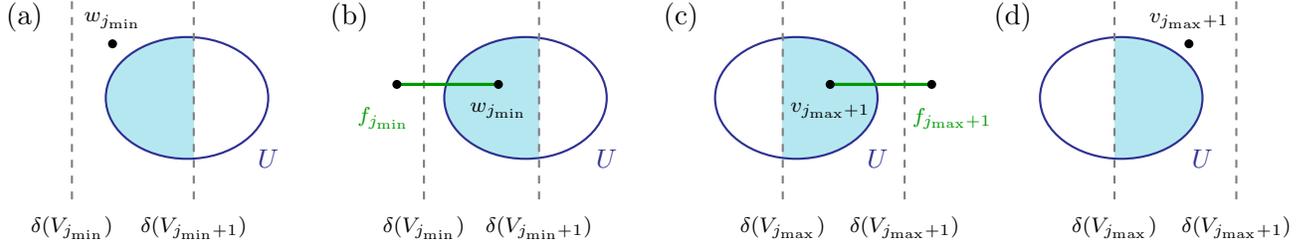
  If $w_{j_{\min}}$ is not contained in $U$ (Figure \ref{fig:cases}~(a)), 
  the set $(V_{j_{\min}+1}\setminus V_{j_{\min}}) \setminus U$ is nonempty,
  and thus, we have $y^{a_{j_{\min}}}(\delta(U)) \ge 1$. This shows 
  \begin{equation}\label{eq:lower_bound_j_min}
   y^{a_{j_{\min}}}(\delta(U)) + \left|\left\{ w_{j_{\min}} \right\} \cap U \right| \ge 1.
  \end{equation}
  Similarly, if $v_{j_{\max}+1}$ is not contained in $U$ (Figure \ref{fig:cases}~(d)), 
  we have $y^{a_{j_{\max}}}(\delta(U)) \ge 1$. This shows 
  \begin{equation}\label{eq:lower_bound_j_max}
   y^{a_{j_{\max}}}(\delta(U)) + \left|\left\{ v_{j_{\max}+1} \right\} \cap U \right| \ge 1.
  \end{equation}
   If $\left|\left\{ w_{j_{\min}} \right\} \cap U \right| =1$, we have $j_{\min} \not = 0$ and 
   $\chi^{f_{j_{\min}}}(\delta(U))=1$ (Figure \ref{fig:cases}~(b)). 
   If $\left|\left\{ v_{j_{\max}+1} \right\} \cap U \right|=1$, we have 
   $j_{\max} < m$ and $\chi^{f_{j_{\max}+1}}(\delta(U))=1$ (Figure \ref{fig:cases}~(c)).
   As we have $j_{\min} \le j_{\max} < j_{\max} + 1$ the edges $f_{j_{\min}}$ (for $j_{\min} >0$) and 
   $f_{j_{\max}+1}$ (for $j_{\max} < m$) are distinct edges. 
   Thus, unless  $j_{\max} = j_{\min}$ and 
   \[\left|\left\{ w_{j_{\min}} \right\} \cap U \right| =
   \left|\left\{ v_{j_{\max}+1} \right\} \cap U \right|  =0,\]
   the inequalities \eqref{eq:lower_bound_j_min} and \eqref{eq:lower_bound_j_max}
   imply $y'(\delta(U)) \ge 2$.  
   
   So it remains to consider the case when $U$ is a subset 
   of $V_{j_{\max}+1}\setminus V_{j_{\max}} = V_{j_{\min}+1}\setminus V_{j_{\min}}$ 
   and contains neither $w_{j_{\min}}$ nor $v_{j_{\max}+1}$. 
   But then \[y'(\delta(U)) \ge y^{a_{j_{\max}}}(\delta(U)) \ge 2.\]   
  \end{proof}

  The next lemma will be used to prove that busy cuts $C$ guessed on levels $l < l$ have a sufficiently large LP value $y(C)$.
  The first part \eqref{eq:busycuts} of the lemma will be applied to guessed busy cuts actually passed to the dynamic 
  program on the current level $l$. (These are the red (densely dotted) busy cuts in Figure \ref{fig:outline}.)
  The second part \eqref{eq:othercuts} of the lemma will be used to show that it is indeed sufficient to pass only 
  guessed busy cuts $\delta(U)$ with $U_1 \cup \{w_1\} \subseteq U \subseteq U_2 \setminus \{v_2\}$, 
  i.e., we do not need to pass the gray (loosely dotted) busy cuts in Figure \ref{fig:outline} to the next level.
  \begin{lemma}\label{lemma:y_lower_bound}
   For every cut $C\in \Bscr$ we have
   \begin{align}\label{eq:busycuts}
    y(C) &\ge 3.
   \end{align}
   For every $U$ with $W_s \subset U \subset V\setminus W_t$ with $s'\notin U$ or $t'\in U$ we have
       \begin{align}\label{eq:othercuts}
    y(\delta(U)) + |\{s'\}\setminus U| + |\{t'\}\cap U| &\ge 3.
   \end{align}
  \end{lemma}
  \begin{proof}
  We first show \eqref{eq:othercuts}.
  For $W_s \subset U \subset V\setminus W_t$  we have by Lemma \ref{lemma:y_lp_solution} that $y(\delta(U))\ge 1$, 
  and if $s',t'\in U$ or $s',t'\notin U$ we have $y(\delta(U))\ge 2$.
  
  To prove \eqref{eq:busycuts} we again use induction on $k-l$.
  For $k=l$ we have $y=x^*$ and the claimed inequality follows from the LP constraints \eqref{eq:dynamic_program_LP}.
  Let now $l<k$.
  We fix a busy cut $C=\delta(U)\in\Bscr$ with $W_s \subset U \subset V\setminus W_t$.
  Note that $s'\in U$ and $t'\notin U$, because busy cuts are $(W_s\cup\{s'\})$-$(W_t\cup\{t'\})$-cuts.
  We will show
  \begin{align}\label{eq:atleastthree} 
   y'(\delta(U)) &\ge  3. 
  \end{align}
  As we have $x^*(C) \ge 3$ by the LP constraints \eqref{eq:dynamic_program_LP} and 
  $y$ is a convex combination of $y'$ and $x^*$, this will complete the proof.
  To show \eqref{eq:atleastthree}, we consider two cases. 
  \begin{figure}
   \begin{center}
    \begin{tikzpicture}[scale=0.83]
      \foreach \s/\x in {9.9/10.5, 
			5.0/5.65} {
	\draw[Blue, thick] (\x,0.5) -- (\x,3.5) ;
	  \draw[Blue, thick, decoration={brace,mirror,raise=1pt, amplitude=6pt},decorate] 
	  (\s-1.5,0.4) -- node[below=6pt] {\scriptsize$U$} (\x,0.4);
      }

      \foreach \s/\l in {4.9/a,
			9.8/b} {
	\node[black] at (\s-1.5,3.9) {\small (\l)};
      }

      \node[darkgreen] at (3.8,1.3) {\small$f_{j}$};
      \node[darkgreen] at (12.4,1.3) {\small$f_{j + 1}$};

      \foreach \x/\j in {9.2/$\delta(V_{j})$, 11.8/$\delta(V_{j+1})$,
			4.5/$\delta(V_{j})$, 6.7/$\delta(V_{j+1})$
			} {
	\draw[thick, dashed, darkgreen] (\x,0.5) -- (\x,3.0) ;
	\node[above] at (\x, 2.8) {\scriptsize \j};
	}

      \draw[ultra thick, darkgreen] (9.8,1.8)-- (12.1,1.8);
      \draw[ultra thick, darkgreen] (4.1,1.8)-- (6.1,1.8);

      \node[circle, fill = black, inner sep = 1.2, outer sep = 1] (v) at (9.8,1.8) {};
      \node [below = 3pt] at (v) {\scriptsize $v_{j+1}$};
      \node[circle, fill = black, inner sep = 1.2, outer sep = 1] (v) at (6.1,1.8) {};
      \node [below = 3pt] at (v) {\scriptsize $w_{j}$};

	\foreach \x in {12.1,4.0} {
	  \node[circle, fill = black, inner sep = 1.2, outer sep = 1] (v) at (\x,1.8) {};
      }

      \end{tikzpicture}
    \caption{Different situations occurring in Case 1 of the proof of Lemma \ref{lemma:y_lower_bound}.
    The picture (a) shows the situation where $w_j \not \in U$. 
    Then $j\not = 0$ and $f_j \in \delta(U)$. 
    The picture (b) shows the situation where $v_{j+1} \in U$. 
    Then $j\not = m $ and $f_{j+1} \in \delta(U)$. \label{fig:U_btw_V_j}}
   \end{center}
  \end{figure}
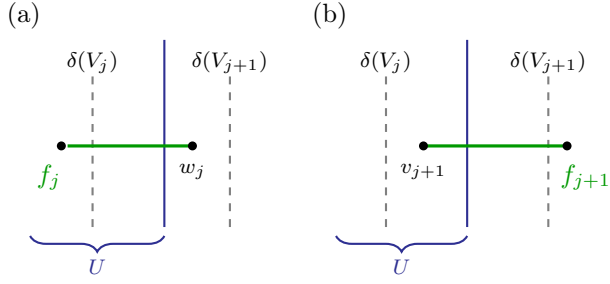 
  \\[2mm]
  \textbf{Case 1:} $V_j \subset U \subset V_{j+1}$ for some $j\in\{0,\dots,m\}$.\\[1mm]  
  We pass $C$ as a busy cut to the next level, i.e. we have $C\in \Bscr^{a_j}$,
  or we have $w_j\not \in U$ or $v_{j+1}\in U$. 
  If $C\in \Bscr^{a_j}$, we apply the induction hypothesis \eqref{eq:busycuts} to the sub-instance corresponding to $a_j$, 
  which implies \eqref{eq:atleastthree} by the definition of $y'$.
  Otherwise we use \eqref{eq:othercuts} and get
  \[ y^{a_j}(C) + |\{w_j\}\setminus U| + |\{v_{j+1}\}\cap U| \ge 3.\] 
  Recall that we have $w_0 = s \in U$ and $v_{m+1} = t \notin U$.
  If $|\{w_j\}\setminus U|=1$, then $j\not=0$ and $\chi^{f_j}(C)=1$.
  If $|\{v_{j+1}\}\cap U|=1$, then $j\not=m$ and $\chi^{f_{j+1}}(C)=1$.
  See Figure \ref{fig:U_btw_V_j}. This implies \eqref{eq:atleastthree} by the definition of $y'$.
  \\[2mm] 
  \textbf{Case 2:} $V_j \subset U \subset V_{j+1}$ holds for no $j\in\{0,\dots,m\}$.\\[1mm] 
  Then the cut $C$ must cross some cut $\delta(V_j)$ with $j\in[m]$, i.e.
  $U\setminus V_j$ and $V_j\setminus U$ are nonempty (see Figure \ref{fig:case_c_crooses_U}).
  Recall that $s'\in V_j \cap U$ and $t' \not \in V_j \cup U$.
  
  \begin{figure}
   \begin{center}
    \begin{tikzpicture}[scale=1.0]
    \begin{scope}
      \clip (-0.5,0.1) rectangle (2,4);
      \draw [fill=SkyBlue, draw = none, opacity=0.4]   (-0.5,0) -- (4,3) -- (4,0) -- cycle;
    \end{scope}
    \begin{scope}
      \clip (2,0) rectangle (4,2.9);
      \draw [fill=SkyBlue, draw = none, opacity=0.4]  (4,3) -- (-0.5,0) -- (0,3) -- cycle;
    \end{scope}
    
    \draw[ultra thick, dashed, darkgreen] (2,0) -- (2,3) ;
    \draw[ultra thick, Blue] (-0.5,0) -- (4,3);
    \node [Blue] (U) at (0.9,1.2) {\small $U$};
    \node [Blue, above] (U) at (4.2,3) {\small $C=\delta(U)$};
    \node[circle, fill = black, inner sep = 1.2, outer sep = 1] (v) at (0.4,2.2) {};
    \node [left = 1pt] at (v) {\small $s'$};
    \node[circle, fill = black, inner sep = 1.2, outer sep = 1] (v) at (3.3, 1.1) {};
    \node [right = 1pt] at (v) {\small $t'$};

    \draw[thick, decoration={brace,mirror,raise=3pt, amplitude=6pt},decorate] (-1,0) -- node[below=8pt] {\small $V_j$} (2,0);
    \end{tikzpicture}

    \caption{Case 2 of the proof of Lemma \ref{lemma:y_lower_bound}, where the busy
    cut $C$ is crossing the cut $\delta(V_j)$, i.e. the two light blue sets 
    $U\setminus V_j$ and $V_j \setminus U$ are nonempty.
    \label{fig:case_c_crooses_U}}
   \end{center}
  \end{figure}
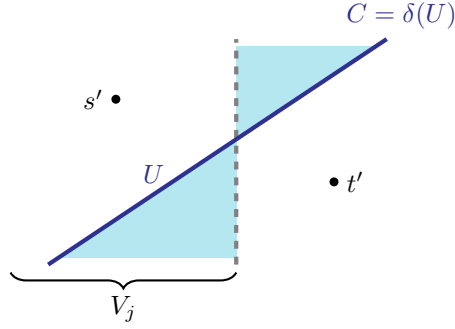
  
  Since neither $s'$ nor $t'$ is contained in $V_j\setminus U$, we have by Lemma \ref{lemma:y_lp_solution}
  $$y'(\delta(V_j\setminus U)) \ge 2.$$ 
  Similarly neither $s'$ nor $t'$ is contained $U\setminus V_j$ and we have by Lemma \ref{lemma:y_lp_solution} that
  $$y'(\delta(U\setminus V_j)) \ge 2.$$ 
  Now by Lemma \ref{lemma:y_1_for_V_j_cuts}, we have $y'(\delta(V_j))=1$. Hence,
  \begin{align*}
  y'(\delta(U)) + 1 &= y'(\delta(U))+ y'(\delta(V_j)) 
  \ge y'(\delta(V_j\setminus U)) + y'(\delta(U\setminus V_j)) 
  \ge 4. 
  \end{align*} 
  This shows \eqref{eq:atleastthree}.
  \end{proof}
  
  We now fix the constants $\lambda_1, \dots,\lambda_k$. We set the scaling constant $\Lambda$ to be 
  $\Lambda := 2^{k+1} -3$. 
  For $l\in[k]$ we set 
  \[ \lambda_l:= \frac{2^{k-l+1}-1}{\Lambda}. \]
  Let $0<\epsilon \le \frac{1}{2}$. 
  We choose the recursion depth $k$ to be 
  \[ k:= \left\lceil \log_2\left(1/\epsilon\right) \right\rceil. \]
  Then we have $k \ge \log_2\left(\frac{3}{2}+\frac{1}{4\epsilon}\right)$
  and thus,
  \begin{align*}
    \lambda_1 =\ &\frac{2^k -1}{\Lambda} = \frac{2^k -1}{2^{k+1} -3} 
  =  \frac{1}{2} + \frac{1/2}{2^{k+1} -3} 
  \le\ \frac{1}{2} + \frac{1}{4 \cdot \left(\frac{3}{2}+\frac{1}{4\epsilon}\right) -6} = \frac{1}{2} + \epsilon.
  \end{align*}
  
  Now we prove that every cut $C$ with a ``small'' LP value $y(C)$ is a guessed lonely cut.
  (Using Lemma \ref{lemma:T_cuts}, we will get that the guessed lonely cuts are no $T$-cuts and thus these cuts
  are not relevant for showing that we obtain a feasible parity correction vector.)
  
  \begin{lemma}\label{lemma:small_cuts_are_in_Lscr}
   If $y(C) < 2 - \frac{1}{\Lambda\cdot \lambda_l}$ for some $(W_s\cup\{s'\})$-$(W_t\cup\{t'\})$-cut $C$, 
   then $C \in \Lscr$.
  \end{lemma}
  \begin{proof} 
  If $l=k$, we have $y(C)\ge 1=2-\frac{1}{\Lambda\cdot\lambda_k}$ by Lemma \ref{lemma:y_lp_solution}.
  Let now $l<k$.

  Let $C=\delta(U)$ with $W_s\cup\{s'\} \subseteq U \subseteq V\setminus(W_t\cup\{t'\})$ and
  $y(C) < 2- \frac{1}{\Lambda\cdot\lambda_l}$.
  By Lemma \ref{lemma:y_lp_solution}, the vector $y$ is a feasible solution to the linear program 
  \eqref{eq:lp_from_lemma_dp}. Hence, the set 
  \begin{align*}
  \Nscr_y := \big\{ \delta(U') \mid\ & y(\delta(U')) < 2, W_s \cup \{s'\} \subseteq U' \subseteq V\setminus (W_t\cup\{t'\}) \big\} 
  \end{align*}
  of narrow cuts is a chain (by Proposition \ref{prop:narrow_cuts}).
  By definition of the sets $V_j$, all cuts $\delta(V_j)$ (for $j\in [m]$) are 
  contained in the set $\Nscr$ of narrow cuts of the vector $x^*$. In particular, we have 
  $x^*(\delta(V_j)) < 2$. By Lemma \ref{lemma:y_1_for_V_j_cuts}, we have $y'(\delta(V_j)) =1$.
  As $y$ is a convex combination of $x^*$ and $y'$, this shows $y(\delta(V_j)) < 2$, and thus,
  $\delta(V_j)\in \Nscr_y$ for all $j\in[m]$.
  From this we can conclude that either $C = \delta(V_j)$ for some $j\in [m]$, or
  $V_j \subset U \subset V_{j+1}$ for some $j\in\{0,1,\dots, m\}$.
  
  If $C = \delta(V_j)$ for some $j\in [m]$, 
  we have $C \in \Lscr$ by construction of $\Lscr$.  
  Otherwise, we have $V_j \subset U \subset V_{j+1}$ for some $j\in\{0,1,\dots, m\}$.
  We distinguish two cases.
  
    \textbf{Case 1:} $C\notin\Nscr$ and $w_j\in U$ and $v_{j+1}\notin U$. \\[1mm] 
    If $C\in \Lscr^{a_j}$, we have $C\in \Lscr$.
  Otherwise we have $y^{a_j}(C)\ge  2-\frac{1}{\Lambda \cdot \lambda_{l+1}}$ 
  by induction.
  Moreover, $x^*(C) \ge 2$.  As 
  \[ y = \frac{\lambda_l - \lambda_{l+1}}{\lambda_l} \cdot x^* + \frac{\lambda_{l+1}}{\lambda_l} \cdot y', \]
  this implies 
  \begin{align*}
    y(C) &
    \ge\ \frac{\lambda_l - \lambda_{l+1}}{\lambda_l} \cdot 2
    + \frac{\lambda_{l+1}}{\lambda_l} \cdot \left(  2-\frac{1}{\Lambda \cdot \lambda_{l+1}} \right) \ 
    =\ 2 \ - \frac{1}{\Lambda \cdot \lambda_l}.
  \end{align*}
   \\[2mm] 
  \textbf{Case 2:} $C \in \Nscr$ or $w_j\notin U$ or $v_{j+1}\in U$. \\[1mm] 
  Then $C\in \Bscr^{a_j}$ or $w_j\notin U$ or $v_{j+1}\in U$.
  By Lemma \ref{lemma:y_lower_bound} applied to this call of the dynamic program, we have
  \[y^{a_j}(C) + |\{w_j\}\setminus U| + |\{v_{j+1}\}\cap U| \ge 3.\]
   If $|\{w_j\}\setminus U|=1$, then $j\not=0$ and $\chi^{f_j}(C)=1$.
  If $|\{v_{j+1}\}\cap U|=1$, then $j\not=m$ and $\chi^{f_{j+1}}(C)=1$.
  Thus,
  \[ y'(C) \ge 3.\]
  
  By the LP constraints \eqref{eq:dynamic_program_LP}, we have $x^*(C) \ge 1$, and hence,
  \begin{align*}
  y(C)\ \ge\ & \frac{\lambda_l - \lambda_{l+1}}{\lambda_l} \cdot x^*(C) 
  + \frac{\lambda_{l+1}}{\lambda_l} \cdot  y'(C)\\[1mm]
   \ge\ & \frac{\lambda_l - \lambda_{l+1}}{\lambda_l} + 3\cdot \frac{\lambda_{l+1}}{\lambda_l} \\[1mm]
   =\ & 2 + \frac{2 \cdot \lambda_{l+1} - \lambda_l}{\lambda_l}\\[1mm]
   =\ & 2 + \frac{2\cdot\left( 2^{k-l} -1 \right) -\left( 2^{k-l+1} -1 \right)}{\Lambda \cdot \lambda_l} \\[1mm]
   =\ & 2 - \frac{1}{\Lambda \cdot \lambda_l}.
  \end{align*} 

  \end{proof}

  \section{The approximation ratio $\mathbf{\frac{3}{2} + \epsilon}$}\label{section:approx_ratio}
  
  \noindent In this section we prove the approximation ratio of $\frac{3}{2} + \epsilon$ for any fixed $\epsilon > 0$.
  Let $S^*$ be the spanning tree, $y^*\in \mathbb{R}^E$ the parity correction vector, and $\Lscr^*$ the chain of cuts 
  returned by the dynamic program with input $W_s =W_t = \emptyset$, $s'=s$, $t'=t$, 
  $\Bscr = \emptyset$, and level $l=1$.

  \begin{lemma}\label{lemma:bound_cost_S_y}
   If OPT denotes the cost of an optimum $s$-$t$-tour, we have 
   \[ c(S^*) + \lambda_1 \cdot c(y^*) \le \left(\frac{3}{2}+\epsilon \right) \cdot \text{OPT}. \]
  \end{lemma}
  \begin{proof} 
  The input of the dynamic program computing $S^*$ and $y^*$ is consistent with any 
  $s$-$t$-tour, in particular with an optimum $s$-$t$-tour $H$
  Thus, we get from Lemma \ref{lemma:bound_on_cost} that
  \[ c(S^*) + \lambda_1 \cdot c(y^*) \le \left(1 + \lambda_1 \right) \cdot c(H). \]
  By the choice of $k$ we have
  \[ 1 + \lambda_1 \le 1 + \frac{1}{2} + \epsilon = \frac{3}{2}+\epsilon, \]
  implying
   \[ c(S^*) + \lambda_1 \cdot c(y^*) \le \left(\frac{3}{2}+\epsilon \right) \cdot \text{OPT}. \]

  \end{proof}
  \bigskip

  \begin{lemma}\label{lemma:y_in_T_join_polytope}
   For \[T=\{ v\in V \mid |\delta(v)\cap S^*| \text{ odd}\} \triangle \{s\}\triangle\{t\}\] the vector 
   $\lambda_1 \cdot y^*$ is contained in the $T$-join polyhedron 
   \[ \{ x \in \mathbb{R}^E_{\ge 0} \mid x(\delta(U)) \ge 1 
   \text { for } |U\cap T|\text{ odd, }\emptyset \not = U \subset V\}.\]
  \end{lemma}
  \begin{proof} 
  From Lemma \ref{lemma:T_cuts} we get that $|S^* \cap C| = 1$ for every cut $C\in \Lscr^*$.
  Moreover, we have that all cuts $C\in \Lscr^*$ are $s$-$t$-cuts. 
  Thus, none of the cuts in $\Lscr^*$ is a $T$-cut, 
  i.e. we have $|U\cap T|$ even for every cut $\delta(U) \in \Lscr^*$.
  Hence, it suffices to show $y^*(C) \ge 1$ for all cuts $C\not \in \Lscr^*$.
  Consider such a cut $C$. By Lemma \ref{lemma:small_cuts_are_in_Lscr}, we have 
  $y^*(C) \ge 2 - \frac{1}{\Lambda \cdot \lambda_1}$. Thus,
  \[ \lambda_1 \cdot y^*(C) \ge 2 \cdot \lambda_1  - \frac{1}{\Lambda} 
   = 2\cdot \frac{2^k -1}{\Lambda}- \frac{1}{\Lambda} = \frac{2^{k+1} -3}{\Lambda} =1.
  \]
 
  \end{proof}
  
  \begin{theorem}
   Let $0<\epsilon \le \frac{1}{2}$. 
   Denote by $p(n,k)$ an upper bound on the time needed to solve a linear program \eqref{eq:dynamic_program_LP} 
   with $|V|= n$ and $|\Bscr| \le k \cdot n$.
   Then there exists a $\left(\frac{3}{2} + \epsilon\right)$-approximation algorithm with runtime 
   $$O\left( n^{6\lceil\log_2(1/\epsilon)\rceil}\cdot p\left(n,\lceil\log_2(1/\epsilon)\rceil\right)\right).$$
  \end{theorem}
  \begin{proof}
  We call the dynamic programming algorithm with level $l=1$, 
  $W_s = \emptyset$, $W_t = \emptyset$, $s'=s$, $t'=t$, 
  and $\Bscr = \emptyset$.
  Let $(V,S^*)$ be the returned spanning tree and $y^*$ the returned parity correction vector.
  We set $T:= \{ v\in V \mid |\delta(v)\cap S^*| \text{ odd}\} \triangle \{s\} \triangle\{t\}$, compute a cheapest $T$-join $J$  
  and an Eulerian trail in $(V,S^* \cupp J)$, and shortcut.
  By Lemma \ref{lemma:y_in_T_join_polytope} the cost $c(S^*)+c(J)$ is at most $c(S^*) + c(y^*)$. 
  By Lemma \ref{lemma:bound_cost_S_y} this is at most $\left(\frac{3}{2} + \epsilon \right) \cdot$ OPT, 
  where OPT denotes the cost of an optimum $s$-$t$-tour.
  
  Calling the dynamic program with level $l=k$ requires solving the linear program \eqref{eq:dynamic_program_LP} once. 
  For $l<k$, the digraph $D$ has at most $n^3$ vertices (because there are at most $n-1$ narrow cuts), and hence
  at most $n^6$ edges. 
  Thus, calling the dynamic program with level $l<k$ requires solving the linear program 
  \eqref{eq:dynamic_program_LP} once, computing the narrow cuts (cf.\ Proposition \ref{prop:narrow_cuts}), 
  and calling at most $n^6$ times the dynamic program with level $l+1$.
  In every recursion step we add only (a subset of the) narrow cuts of the computed LP solution to the set $\Bscr$. 
  As the narrow cuts form a chain, these are at most $n$ cuts. 
  Thus, for the recursion depth $k=\left\lceil \log_2\left(1/\epsilon\right) \right\rceil$
  we have $|\Bscr|\le\left\lceil \log_2\left(1/\epsilon\right)\right\rceil\cdot n$.
  The runtime is dominated by solving one LP and calling the dynamic program at most $n^6$ times recursively.
  If we denote by $t_l$ the maximum runtime of the dynamic program on level $l$ (including recursive calls),
  then $t_k \le p\left(n,\lceil\log_2(1/\epsilon)\rceil\right)$ and 
  $t_l = O\left(p\left(n,\lceil\log_2(1/\epsilon)\rceil\right) + n^6 \cdot t_{l+1} \right)$ for $1\le l < k$.
  By induction on $k-l$, we obtain a runtime of 
  $t_l =  O\left( n^{6(k-l)}\cdot p\left(n,\lceil\log_2(1/\epsilon)\rceil\right)\right)$.
  \end{proof}
  One can improve the $n^{6\lceil\log_2(1/\epsilon)\rceil}$ bound to $n^{4\lceil\log_2(1/\epsilon)\rceil}$
  by observing that there are at most $n^4$ sub-instances of any instance.
  Note that $p(n,k)$ can be chosen as a polynomial because the busy cut constraints can be checked explicitly, 
  and the separation problem for the other cut constraints reduces to $O(n)$ minimum cut computations.
  Hence, we have a polynomial-time algorithm for any fixed $\epsilon>0$.
  
  We remark that we do not need the explicit LP solutions for our algorithm.
  The only properties we use from the LP solutions are the LP value and the set of narrow cuts.
  
\subsection*{Acknowledgements}
We thank the anonymous reviewers for their helpful comments and suggestions.

\end{document}